\documentclass{amsart}
\usepackage{cite}

\usepackage{amsmath}								
\usepackage{amssymb}								
\usepackage{commath}                                

\usepackage{cite}									

\usepackage{nicefrac}                               
\usepackage{pdfsync}                                
\usepackage{units}                                  

\newcommand\myHeight{4.2}

\usepackage[letterpaper,margin=1in]{geometry}

\usepackage[pdftex]{graphicx}
\DeclareGraphicsExtensions{.pdf,.jpeg,.png,.eps}

\usepackage{epstopdf}								

  \usepackage[caption=false,font=footnotesize]{subfig}

\begin{document}
%
\title{Simulation and Testing Results for a Sub-Bottom Imaging Sonar}



\author[D.C. Brown]{Daniel C. Brown}
\author[S.F. Johnson]{Shawn F. Johnson}
\author[C.F. Brownstead]{Cale F. Brownstead}
\address{Applied Research Laboratory \\ State College, PA 16804--0030}
\email{dan.brown@psu.edu, shawn.johnson@psu.edu, cfb102@psu.edu}


\maketitle

\begin{abstract}
The problem of detecting buried unexploded ordnance (UXO) is addressed with a sensor deployed from a shallow-draft surface vessel. This sonar system produces three-dimensional synthetic aperture sonar (SAS) imagery of both surficial and buried UXO across a range of environments. The sensor’s hardware design was based in part upon data created using a hybrid modeling approach that combined results from separate environmental scattering and target scattering models. This hybrid model produced synthetic sensor data where the sensor/environment/target space could be modified to explore the expected operating conditions. The simulated data were also used to adapt a set of existing signal processing algorithms for formation of three-dimensional acoustic imagery.

Recently, the sonar system has been integrated to a test platform, and experiments have been conducted at a trial site in the Foster Joseph Sayers Reservoir near Howard, PA. This test site has been prepared with several buried man-made objects. Initial results show that fully buried targets can be detected.
\end{abstract}


%

\section{Introduction}
The remediation of unexploded ordnance (UXO) is a current environmental problem facing the United States Department of Defense \cite{SERDP:2007a,SERDP:2013a,SERDP:2018a}. UXO can be found in a number of aquatic environments, and over time the ordnance may become buried \cite{SERDP:2014a}. Those environments where the ordnance is near shore are of particular concern, and sensors capable of detailed survey are needed for detecting and localizing UXO in these environments.

The problem of buried UXO detection has been addressed in prior research with sonar imaging systems \cite{Schock:2001a,Schock:2002a,Schock:2005a,Schock:2006a}. One advantage of the sonar sensing modality (over electromagnetic modalities) is that acoustic imaging offers the promise of higher potential area coverage rates and better localization. The prior sonar systems that have addressed the problem of buried UXO imaging have either been towed systems or deployed from unmanned underwater vehicles. This deployment method has limited their operation to waters typically deeper than \unit[5]{m}.

The problem of UXO detection and classification is complicated by the very wide range of potential UXO targets. UXO remediation sites may have ordnance as small as individual bullets whose largest dimension is less than \unit{1}{cm} up to bombs that may exceed \unit[1]{m} in length. Additionally, the ordnance may experience significant biofouling and corrosion after remaining in place for several decades. Finally, man-made clutter is commonly found in near-shore UXO surveys. The range of targets sizes, the variability of the target state, and the presence of clutter requires high-resolution imaging sensors for effective performance.

\begin{figure}[t]
  \centering
  \includegraphics[width=.65\columnwidth]{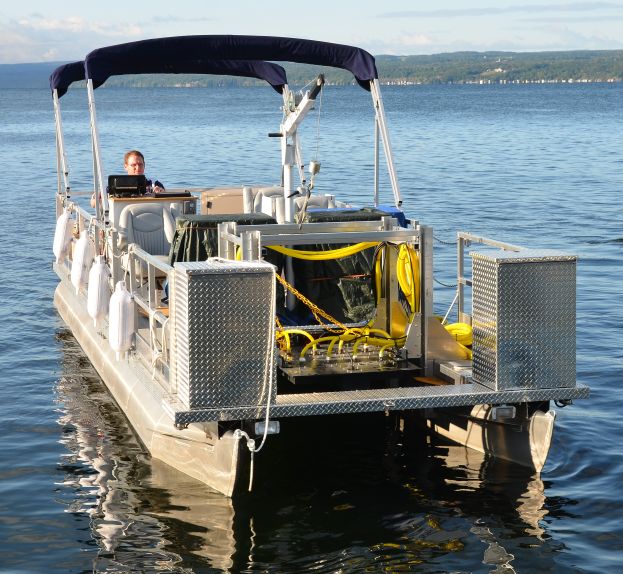}
  \caption{The test platform is a nine-meter pontoon boat. The projector and receive array are shown here mounted in the forward portion of the boat. These components are mounted in a rectangular frame that is lowered into the water during testing.}\label{fig:soundHunter}
\end{figure}

A prototype sensor, called the Sediment Volume Search Sonar (SVSS), has recently been developed to address the shallow-water buried UXO problem. This sonar system uses five discrete projectors and a two-dimensional receive array to create high-resolution three-dimensional imagery through synthetic aperture signal processing. This sensor is integrated in a nine meter pontoon boat, Figure~\ref{fig:soundHunter}, to enable operation with water depths as shallow as \unit[1]{m}.

The sensor's array design is based in part upon a modeling and simulation collaboration with the Applied Physics Laboratory at the University of Washington (APL-UW)). This collaboration produced synthetic sensor data, where the sensor-environment-target space could be modified to explore the expected operating conditions. The approach utilized a pair of models; one for environmental scattering and one for target scattering. The field scattered from the environment is simulated using the Applied Research Laboratory / Penn State University (ARL/PSU) Point-based Sonar Scattering Model (PoSSM) \cite{Brown:2017b}. This model was combined with the APL-UW developed Target in the Environment Response (TIER) model \cite{Kargl:2015a}. PoSSM and TIER both produce calibrated, bistatic, frequency-dependent, element-level, time series suitable for coherent signal processing. Details of the hybrid modeling approach and results are provided in Section~\ref{sec:modeling}.

The test site and experimental results are detailed in Section~\ref{sec:fieldExperimentation}. A prototype SVSS sensor was integrated to the test platform shown in Figure~\ref{fig:soundHunter}. Sensor testing occurred in late 2017 at a trial site in central Pennsylvania, which was prepared with several man-made objects buried up to \unit[20]{cm} depth.

\section{Sensor Modeling} \label{sec:modeling}

\subsection{PoSSM \& TIER Modeling Approach}
Prior to hardware fabrication, design alternatives for a sub-bottom imaging sensor were evaluated using a hybrid modeling approach. This approach combined target scattering results from TIER with environmental scattering results from PoSSM. The integration of PoSSM and TIER provides a high-speed, coherent model for the simulation of the field scattered from buried UXO and the surrounding environment. This hybrid model enabled analysis of sensor performance as a function of array design, target properties and environmental properties.

PoSSM is a point-based scattering model, where the individual scattering amplitudes are calculated using deterministic physical models as well as a stochastic scale factor. Figures~\ref{fig:compIntVol21}~and~\ref{fig:compIntVol09} show examples of scattering points with shading indicating the composite levels for very fine silt and medium sand, respectively. Geoacoustic properties for each sediment are provided in Table~\ref{tab:hfevaSediments}. These composite levels are calculated by combining models for propagation loss, sensor directivity, and diffuse interface and volume scattering. The coherent component of the scattered field is modeled using Eckhart's approximation to the reflection coefficient \cite{Eckart:1953a}. Figures~\ref{fig:possmTimeSeries21}~and~\ref{fig:possmTimeSeries09} show the time series resulting from the scatterer arrangement in the associated figures to the left. These plots show the individual time series components as well as the envelope of the composite result. The PoSSM model has a mathematically simple form which allows easy implementation and efficient computation. Results of this model have been compared to the sonar equation, and both the mean field and mean square field show good agreement. The spatial coherence of the scattered field was also shown to agree with the van Cittert-Zernike theorem for an idealized environment.

\begin{table}[t]
  \centering
  \caption{Sediment properties for the two bottom types used for the numerical modeling with the TIER and PoSSM simulations. Values are taken from \cite{APL-UW:1994a}, and the attenuation coefficients are provided at \unit[20]{kHz}.} \label{tab:hfevaSediments}
    \begin{tabular}{l|c|c}
    \hline
    \textbf{Sediment} & \textbf{Medium} & \textbf{Very Fine} \\
    \textbf{Property} & \textbf{Sand} & \textbf{Silt} \\
    \hline
    Density $[\textrm{kg/m}^{3}]$ & 1845  & 1147 \\
    Sound Speed [m/s] & 1767  & 1476 \\
    Attenuation Coefficient [dB/m] & 10.0  & 1.4 \\
    Spectral Strength $[\textrm{m}^{(4-\gamma)}]$  & 1.410E-04 & 1.638E-05 \\
    Spectral Exponent [unitless] & 3.25  & 3.25 \\
    Vol. Scat. Strength [dB/m] & -20.0 & -28.6 \\
    \hline
    \end{tabular}%

\end{table}%

\renewcommand\myHeight{4.9}

\begin{figure}[t]
  \centering
  \subfloat[]{\label{fig:compIntVol21}\includegraphics[height= \myHeight cm]{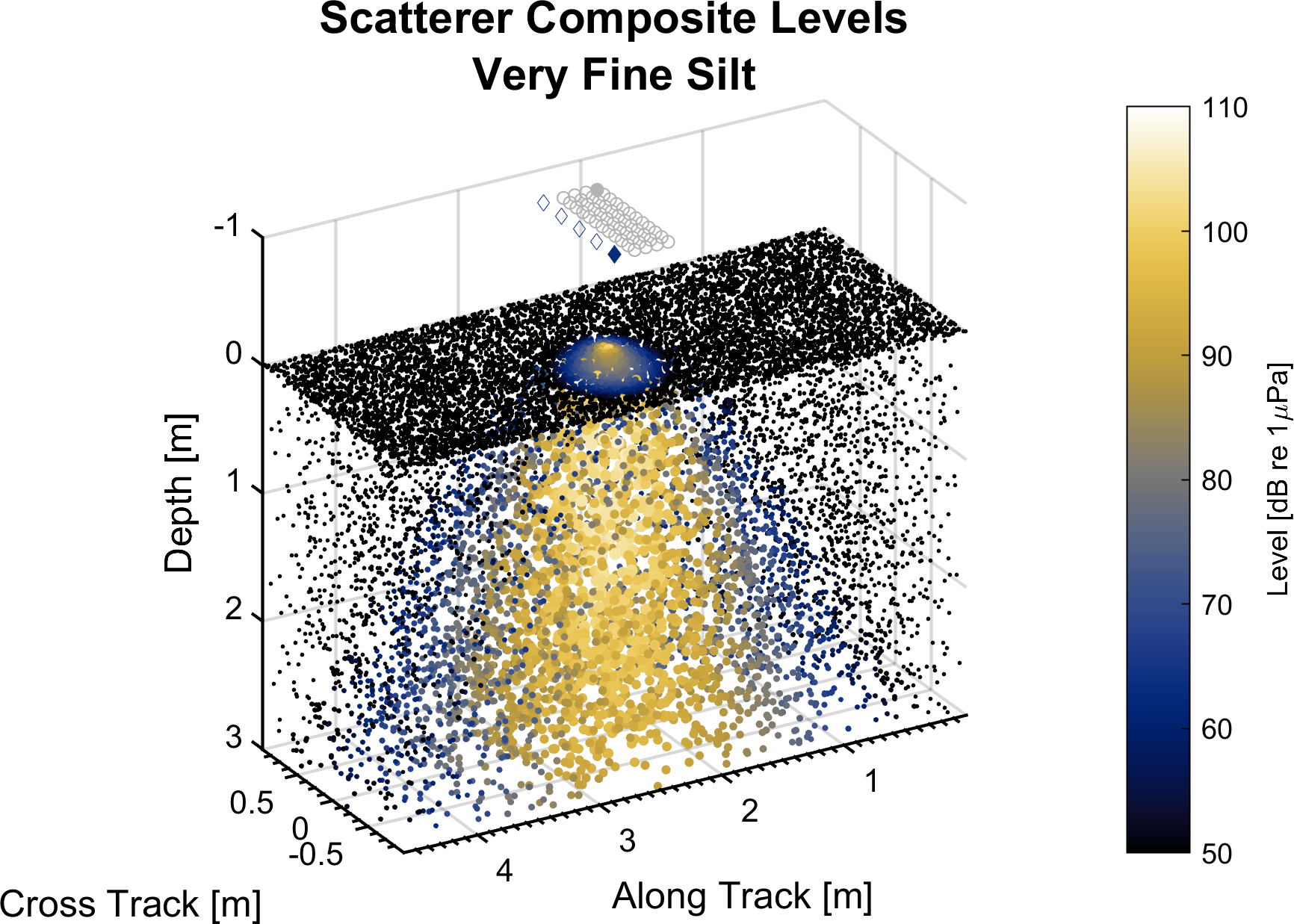}}
  \hspace{5mm}
  \subfloat[]{\label{fig:possmTimeSeries21}\includegraphics[height= \myHeight cm]{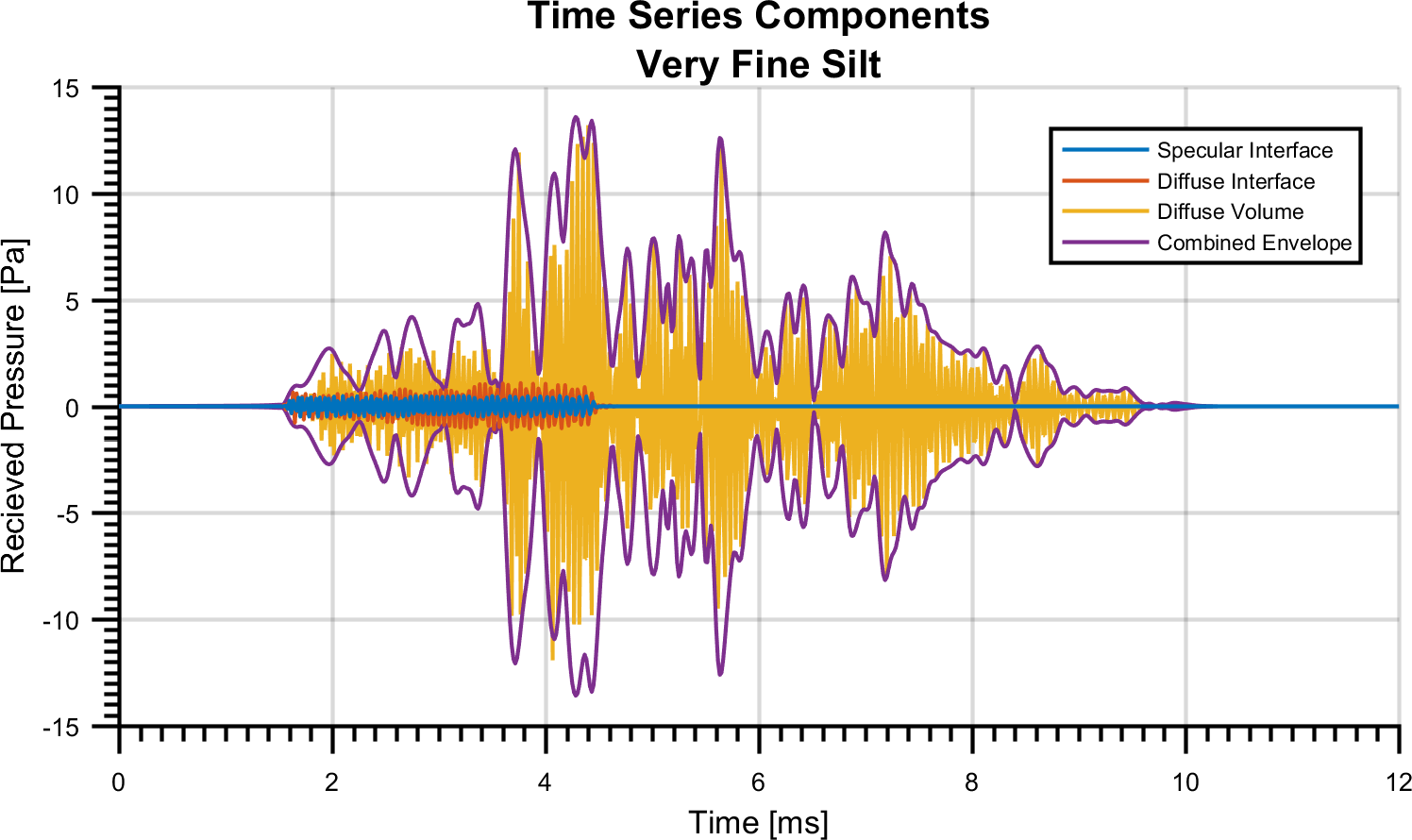}}
  \vspace{1mm}
  \subfloat[]{\label{fig:compIntVol09}\includegraphics[height= \myHeight cm]{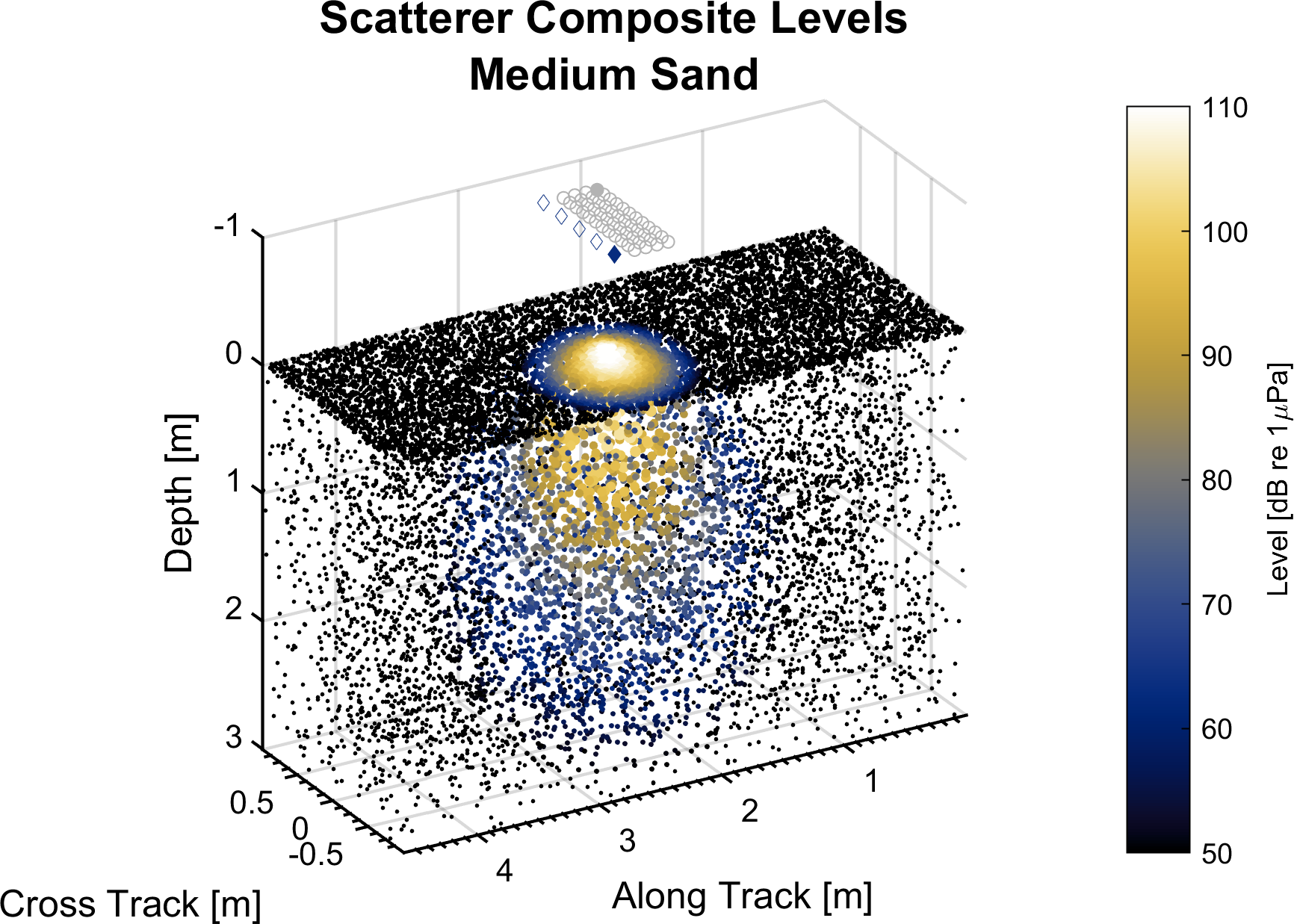}}
  \hspace{5mm}
  \subfloat[]{\label{fig:possmTimeSeries09}\includegraphics[height= \myHeight cm]{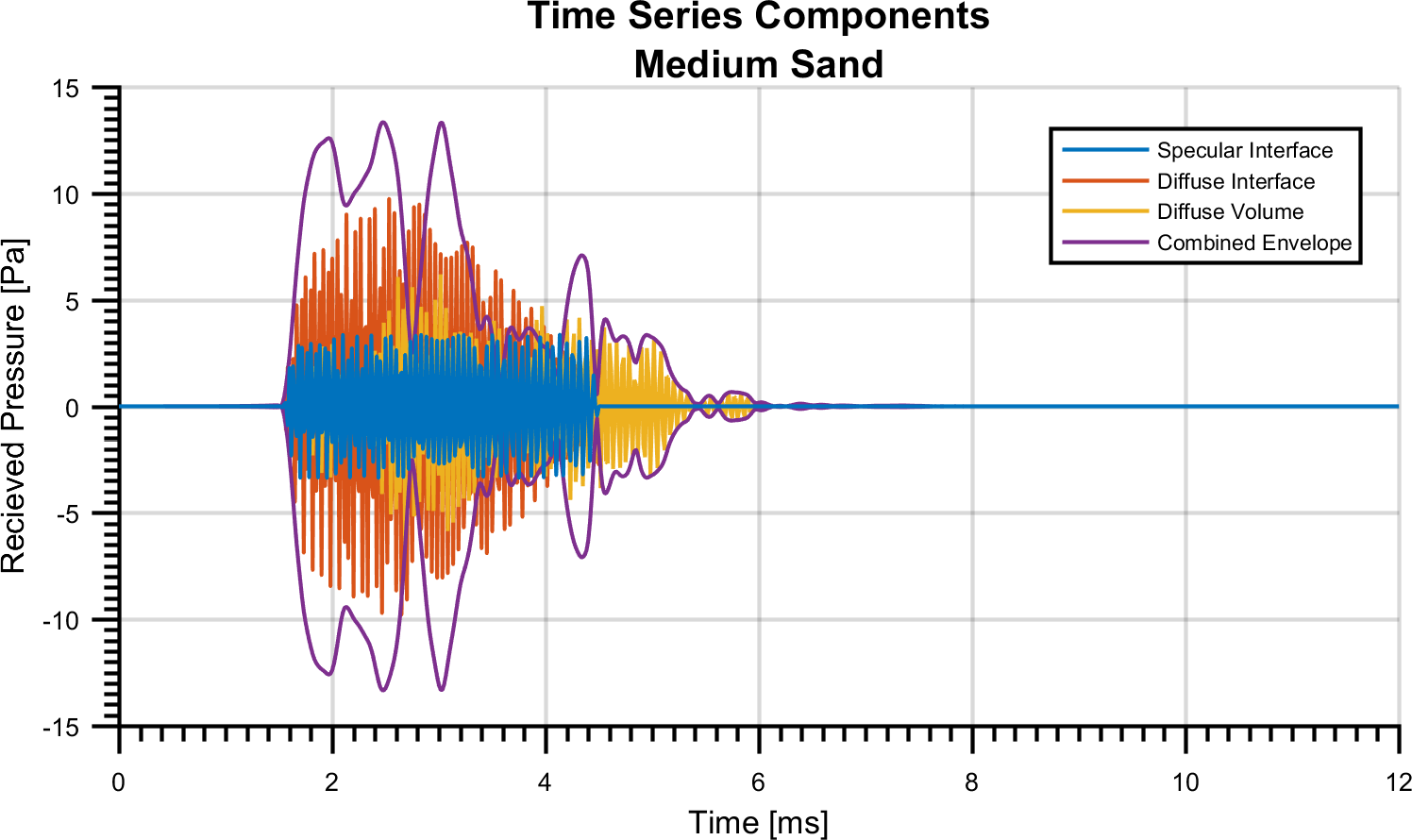}}
  \caption{Diffuse interface and volume scatterer composite levels are shown for a single ping, between a single transmit-receive pair at the extreme corners of the sonar. The composite scatterer level includes the source level and effects of transmit and receive beampatterns, model-based scattering strength, attenuation, spreading loss, and for volume scatterers the two-way transmission through the interface.}\label{fig:compIntVol}
\end{figure}

APL-UW's TIER model provides a coherent simulation for target scattering that is needed in order to evaluate a sensor's ability to image buried UXO. TIER's propagation model assumes the water is a non-attenuating fluid, while the sediment is treated as an attenuating fluid, with properties given in Table~\ref{tab:hfevaSediments}. For simple shapes, TIER uses analytic form functions to model target scattering. For more complex shapes, a finite element model is employed to provide a lookup table. Both target scattering models include the full elastic field scattered from the targets.

Two noise sources were considered in the hybrid model. Ambient noise, which was assumed to be omnidirectional, was included at a level matching sea state 3 levels reported for deep water spectra in Urick \cite{Urick:1983a}. The other noise source is multipath interference. Multipath is known to limit sonar image quality for stripmap SAS \cite{Hansen:2011a}, and the same mechanisms will interfere with a sub-bottom imaging sensor. In the definition employed here, multipath noise consists of all acoustic signals generated by the sonar transmitter that propagate out and return at the same time instant as the signal of interest. The level of the multipath returns are proportional to the transmit level; therefore, increasing source level does not improve the signal-to-noise ratio in multipath limited conditions.

A model for multipath interference for mid- and high-frequency SAS imagery has been proposed by Lowe and Brown \cite{Lowe:2012a}. In this model, the level of multipath interference is modeled using a simple ray-based approach. Each boundary interaction accumulates a loss factor along with an additional loss associated with the directivity of the transducers. Along each ray path, the losses are accumulated as a function of ray path length. If it is assumed that the air-water interface is perfectly flat, then the multipath model proposed by Lowe and Brown may be easily merged with PoSSM by exploiting the method of images \cite{Pierce:1991a}. In this approach, multipath reflections from a flat planar interface are simulated using an image source whose position is reflected about the boundary. Finally, experimental observations have shown that multipath interference in high frequency SAS imagery is highest at low sea surface roughness \cite{Hansen:2011a}. This is because the low roughness interface minimizes loss at the air-water boundary. Thus, the assumption of a flat air-water interface provides an upper bound on multipath interference.

The modeled sonar array consists of a 32-channel receive array with the elements arranged in a fully-populated two-dimensional grid with 4 channels in the along-track direction and 8 channels in the cross-track direction. The receiver elements are separated by \unit[9.1]{cm} in each direction. The transmit array consists of five discrete projectors, which are positioned forward of the receive array, separated by \unit[22.9]{cm} in the cross-track direction. The sonar is simulated with each projector transmitting independently, so that sound is only produced from one projector for each ping. At each ping location, the PoSSM and TIER models are executed for each of the receivers to produce 48 time series per ping. The sonar advances to the next location for a total of 51 ping locations. For a complete simulation of a sonar scan with a single transmitter, 2448 individual ping locations are simulated for the 5 transmitter SVSS configuration, 12,240 individual PoSSM+TIER simulations were produced. This process repeats for a range of sonar / target configurations.

\subsection{Modeling Results}
Several targets were modeled in the sensor analysis, but this section will focus on the results from a \unit[61]{cm} long \unit[31.5]{cm} diameter solid aluminum cylinder. In the simulation, the sensor operates at an altitude of \unit[2]{m} above the sediment-water interface and the total water depth is \unit[2.5]{m}. The RMS roughness of the interface is assumed to be \unit[1]{cm} for the purposes of calculating the coherent reflection coefficient. The target is buried at \unit[1]{m} depth beneath the sediment water interface and directly beneath the sensor's track. Finally, all beamforming is conducted assuming perfect knowledge of the sediment sound speed so that the effects of refraction are removed.

\begin{figure}[t]
    \subfloat[]{\label{fig:simSilt}\includegraphics[width= .48\columnwidth]{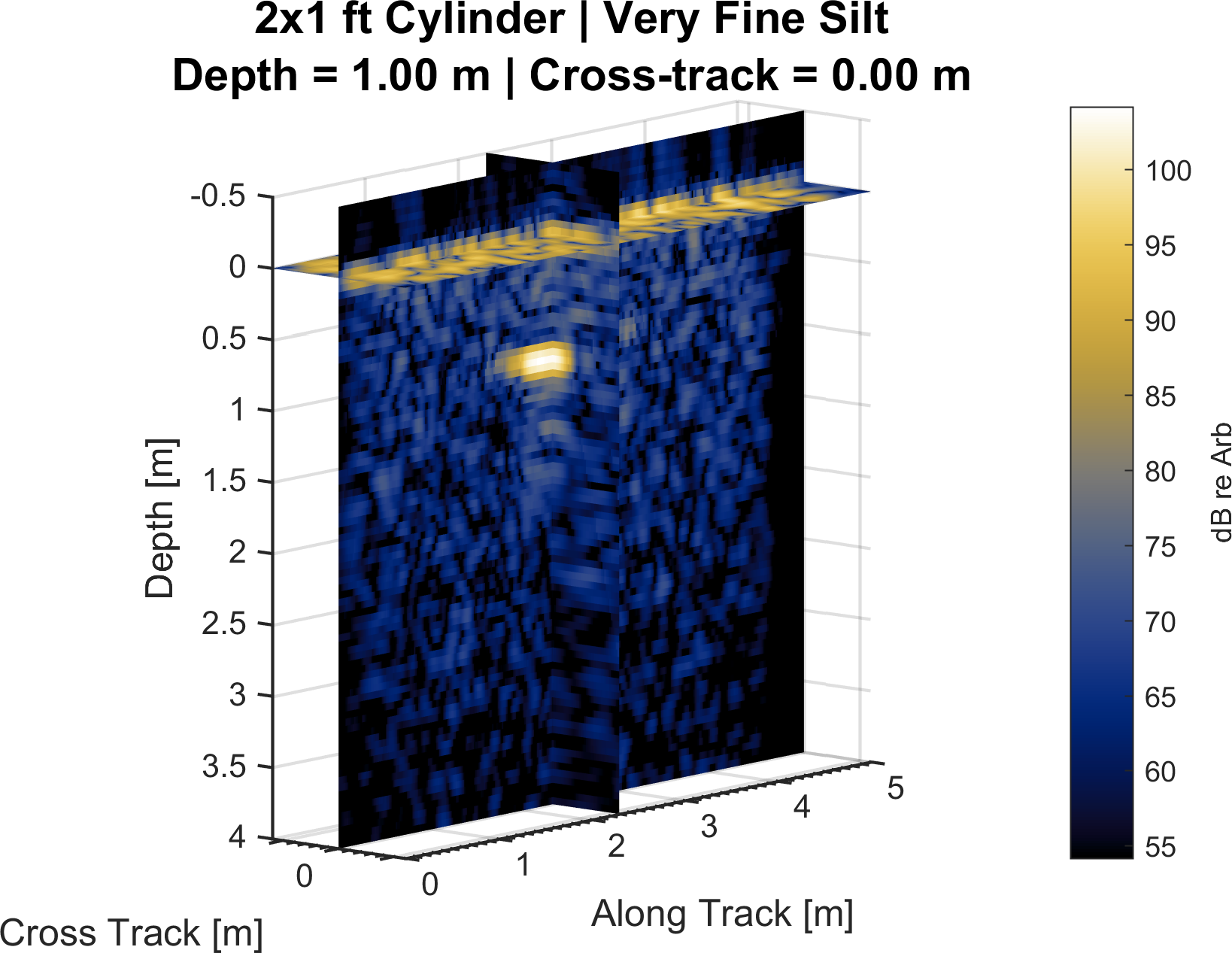}}
    \hspace{1mm}
    \subfloat[]{\label{fig:simSand}\includegraphics[width= .48\columnwidth]{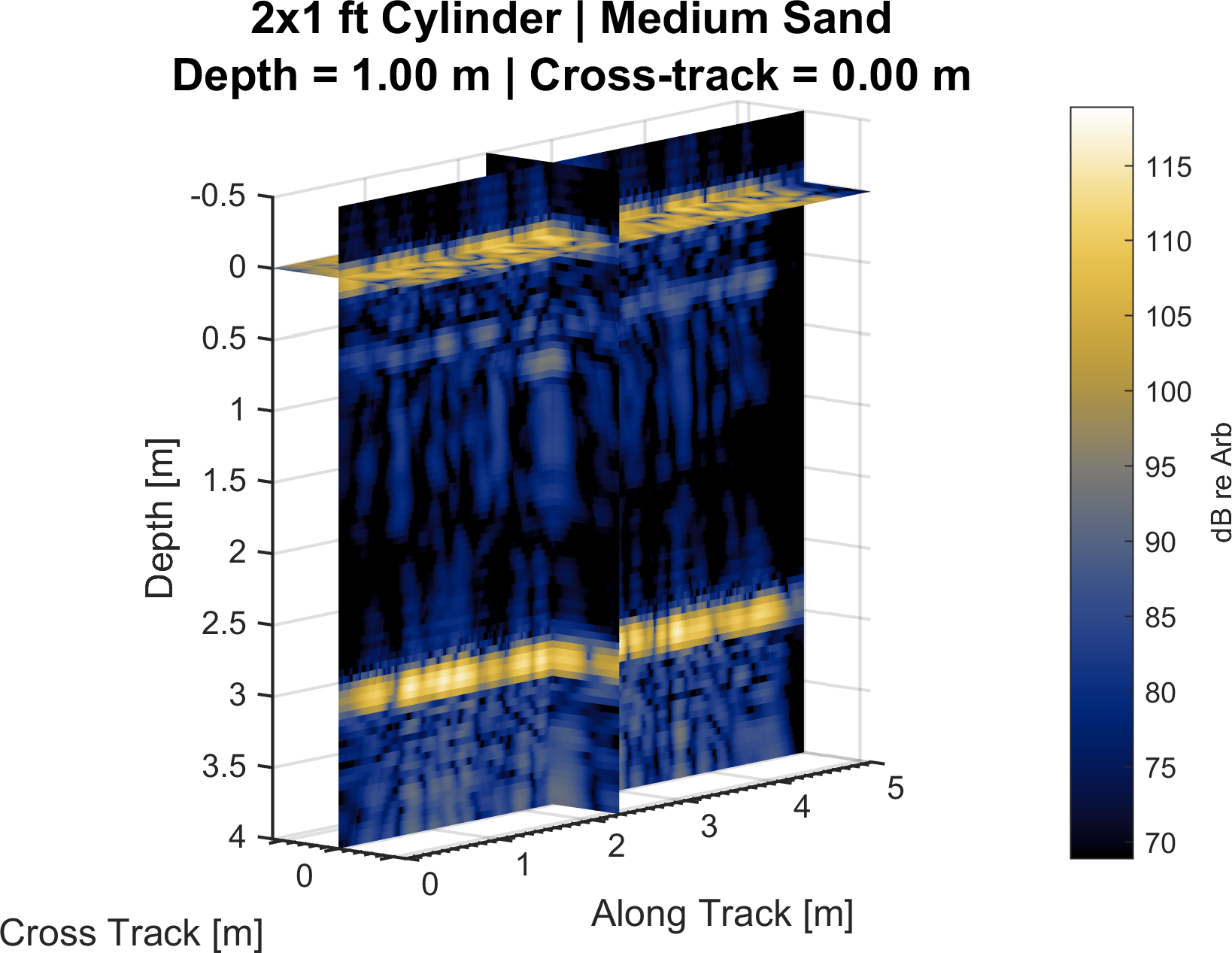}}
    \caption{The large cylinder is shown at a burial depth of \unit[1.0]{m} in the silt and sand sediments.}\label{fig:simTargetPair}
\end{figure}

The beamformed results of the simulations for the buried target are shown in Figure~\ref{fig:simTargetPair}. In each of these images, three planar slices through the peak of the target response are shown in a three-dimensional view. Figure~\ref{fig:simSilt} shows the target buried in the very fine silt sediment. Because of the relatively low sediment attenuation and low volume scattering strength, the target is clearly distinguishable from the background.

Figure~\ref{fig:simSand} shows the same target for a medium sand sediment. Due to the sediment attenuation, a depth varying gain of \unitfrac[10]{dB}{m} is applied to this data for visualization purposes. The combination of an increased scattering strength and attenuation has increased the multipath interference level for this environment. The first-order multipath rays are seen at a sediment depth of \unit[0.6]{m}. The second-order multipath ray appears at \unit[2.9]{m} sediment depth. The target is visible at \unit[1]{m} depth, but the signal excess is less than that found for the very fine silt case. A detailed look at the data shows the target is \unit[10-11]{dB} above the background in this environment. The target is detectable; however, it is apparent that if the target was buried at a slightly shallower depth it may have been obscured by the multipath interference.

\section{Field Experimentation and Results} \label{sec:fieldExperimentation}

\subsection{Test Site Development}
\begin{figure}[t]
  \centering
  \subfloat[Winter -- Reduced Pool Height]{\label{fig:sayersZoom2006}\includegraphics[width= .45\columnwidth]{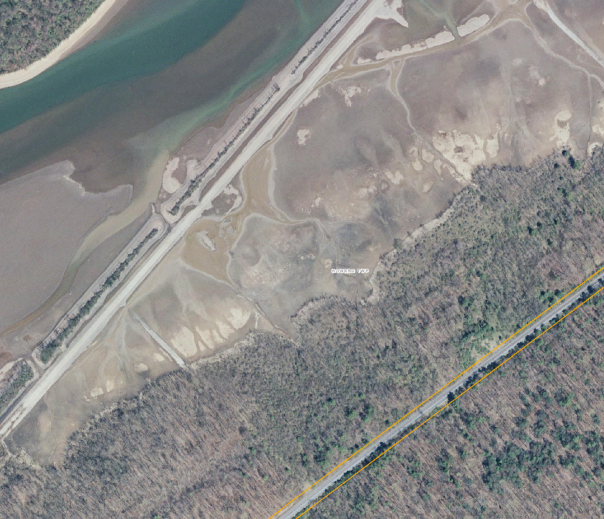}}
  \vspace{1mm}
  \subfloat[Spring/Summer/Fall -- Full Pool Height]{\label{fig:sayersZoom2014}\includegraphics[width= .45\columnwidth]{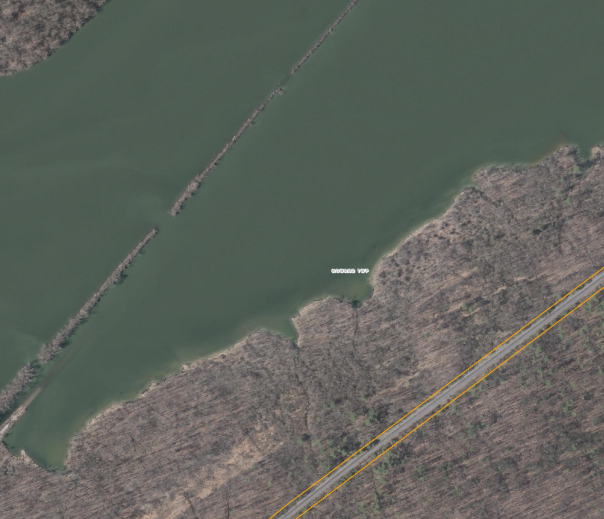}}
  \caption{This figure provides a view of the region of the Foster Joseph Sayers Reservoir where a test site is installed. The winter reduction of the pool height provides a unique opportunity for installing a target field with ground truth on target positions and burial depths. Imagery provided by the Pennsylvania Spatial Data Access and used with permission \cite{PSDA:2018a}.} \label{fig:sayersZoom}
\end{figure}

Foster Joseph Sayers Reservoir, near Howard, PA approximately 20 miles NNE of ARL/PSU, was created in 1971 by the United States Army Corps of Engineers for flood control. This lake covers roughly 1,700 acres and is eight miles in length. During the winter months, typically December to April, the pool height is lowered to accommodate springtime snowpack runoff. This procedure annually exposes a portion of the lakebed in the late winter. The test site was chosen due to its close proximity to ARL/PSU, and because the winter lowered pool height provided a unique opportunity for establishing an accurate ground truthed test bed. At low pool height during March 2017, two testing sites were prepared, with sonar testing being conducted during the summer of 2017 at full pool height. Figure~\ref{fig:sayersZoom2006} shows the lake level near the lowest level achieved during maximum winter drawdown. Figure~\ref{fig:sayersZoom2014} shows the lake at the full summer pool level. Additionally, the remnants of a road on the southwest edge of this field are still visible in the bottom left corner of Figure~\ref{fig:sayersZoom2006}. This road provides a firm bottom for staging equipment during test field installation.

An initial set of targets were deployed in early March of 2017. The installation process included one day of staging the targets and equipment along the exposed roadbed near the test sites. This staging exercise was followed by two days of target installation. A partial list of the deployed targets and their properties is provided in Table~\ref{tab:sayersTargets}. The target strengths listed in this table are high frequency ($ka\gg1$) approximations taken from Urick\cite[Table 9.1]{Urick:1983a}. These targets were selected to provide a range of sizes for evaluation of the SVSS while also being small enough to be manually transported 1.2 miles from the shore of the lake.

\begin{table}[t]
  \centering
  \caption{Three target types were installed at the Foster Joseph Sayers Reservoir test site. The broadside target strength approximation is taken from Urick \cite{Urick:1983a}.}\label{tab:sayersTargets}%
    \begin{tabular}{l|c|c|c}
    \hline
    \textbf{Target} & \textbf{Target} & \textbf{Target} & \textbf{Approx. Broadside} \\
    \textbf{Name} & \textbf{Length} & \textbf{Diameter} & \textbf{Target Strength} \\
          & \textbf{[cm]} & \textbf{[cm]} & \textbf{[dB]} \\
    \hline
    Shot put - 12 lb. & N/A   & 10.2  & -31.9 \\
    Solid Al Short Cylinder & 30.5  & 15.2  & -11.9 \\
    Solid Al Long Cylinder & 61.0  & 15.2  & -5.9 \\
    \hline
    \end{tabular}%

\end{table}%

\subsection{SVSS Sensor and Signal Processing}

\begin{figure}[tb]
  \centering
  \includegraphics[width=.45\columnwidth]{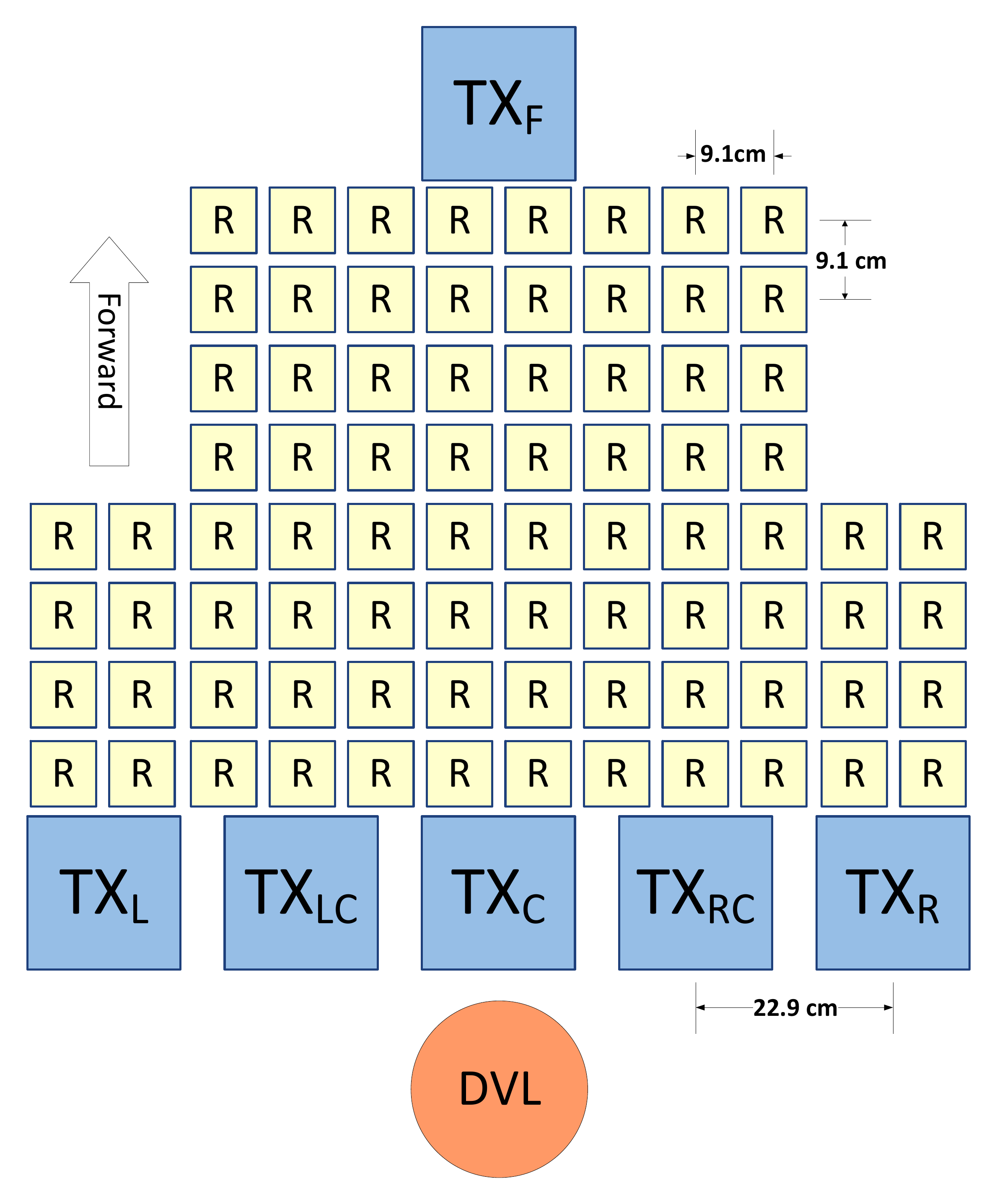}
  \caption{A schematic diagram of the SVSS array configuration is shown. The array consists of 80 receive channels, which are labeled R, and six projectors, which are labeled $\textrm{TX}_\textrm{XX}$. The SVSS can also utilize a Doppler Velocity Log (DVL) which is mounted aft of the sonar array.} \label{fig:arraySchematic}
\end{figure}

The SVSS receive array is made up of a series of eight-channel hydrophone modules. These modules are designed so that when flush mounted they create arrays with a \unit[9.1]{cm} center-to-center spacing. In the modeling and simulation phase, the array consisted of six of these eight-channel modules that were arranged in a 4x12 channel grid. Four new hydrophone modules were fabricated prior to field testing, and the array was expanded to add a 4x8 channel section forward of the originally planned 4x12 channel section. This layout is seen in Figure~\ref{fig:arraySchematic}. In this arrangement, the sonar array is roughly \unit[73]{cm} in the along-track direction and \unit[1.1]{m} in the cross-track direction.

A backprojection beamformer is utilized to post-process SVSS data, creating imagery with voxels (i.e. three-dimensional pixels) that are \unit[2]{cm} along each dimension. For the SVSS data presented here, the image output size is \unit[2]{m} cross-track, \unit[15]{m} along-track, and \unit[2]{m} depth. Using a \unit[2]{cm} resolution, the output imagery is a 100x750x100 data cube. Because individual tracks are all greater than \unit[15]{m} along-track, each track is beamformed into a series of images of this length. This along-track length is a user-selected parameter and may be modified depending on the test conditions.

The detection of targets in the volumetric imagery is complicated by the reflectivity of the silt/clay boundary. Some points along the boundary actually exceed the peak scattering level observed for the cylindrical target. The use of a linear intensity limits visual interpretation of returns from the deeper sediment. The strong reflection from the silt/clay boundary, combined with the spreading loss and sediment attenuation, reduces the strength of the sub-bottom return to the point where it is not visible with this image. Mitigation of the wide dynamic range present in SVSS imagery requires background estimation (and normalization) to make the imagery interpretable. This type of processing is commonly used within the sonar machine learning community \cite{Dobeck:2010a,Williams:2018a}. For data presented here, the background is estimated using a median filter with a three-dimensional kernel whose dimensions are \unit[0.2]{m} cross-track by \unit[1.2]{m} along-track by \unit[0.1]{m} depth. These filter dimensions were determined heuristically through analysis of image interpretability. After the raw data are normalized by the background estimate, they are further processed with a dynamic range compression algorithm. This algorithm applies a combination of logarithmic mapping and nonlinear intensity mapping operators to further compress the most extreme sample values to aid visual inspection.

\subsection{Results}
\begin{figure*}[!t]
  \centering
  \includegraphics[width=.95\textwidth]{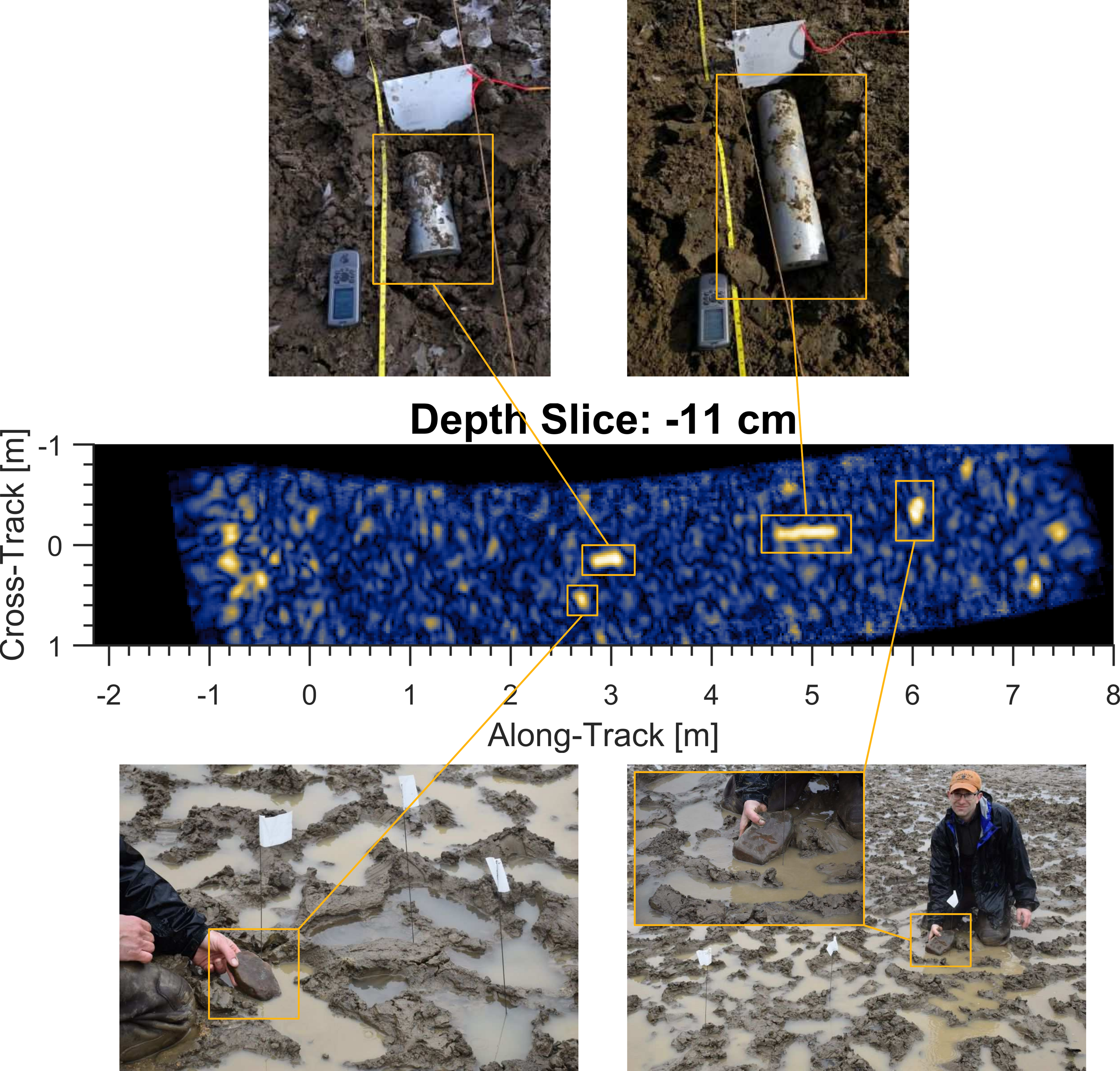}
  \caption{Two clutter targets were identified in the depth slice shown in the middle figure. During the installation of a second target field in March 2018, the clutter objects were determined to be the pair of rocks shown in the bottom photographs.} \label{fig:rocksAndSlice}
\end{figure*}

The depth slice shown in Figure~\ref{fig:rocksAndSlice} shows a pair of cylindrical targets at \unit[3]{m} and \unit[5]{m} along-track. There are also several additional returns that are nearly the same scattered level in the imagery. In particular, there is a bright return at \unit[2.7]{m} along track and \unit[0.5]{m} cross-track, and another bright return at \unit[6]{m} along track and \unit[-0.3]{m} cross-track. During the installation of the second test field in March of 2018, the ends of two cylindrical targets were located and marked with flags. Using the sonar imagery, the locations of the two clutter objects relative to the two cylindrical targets were calculated. Upon placing flags at both calculated positions, the two flat rocks shown in the lower frames of Figure~\ref{fig:rocksAndSlice} were found and excavated. Each rock was found with a flat face oriented upward, which is the likely cause for the relatively strong acoustic return. This post-acoustic-survey localization and recovery of modest sized buried objects serves as a demonstration of the sensor's potential for detection and localization of relatively small objects.

\begin{figure*}[!t]
  \centering
  \includegraphics[width=.9\textwidth]{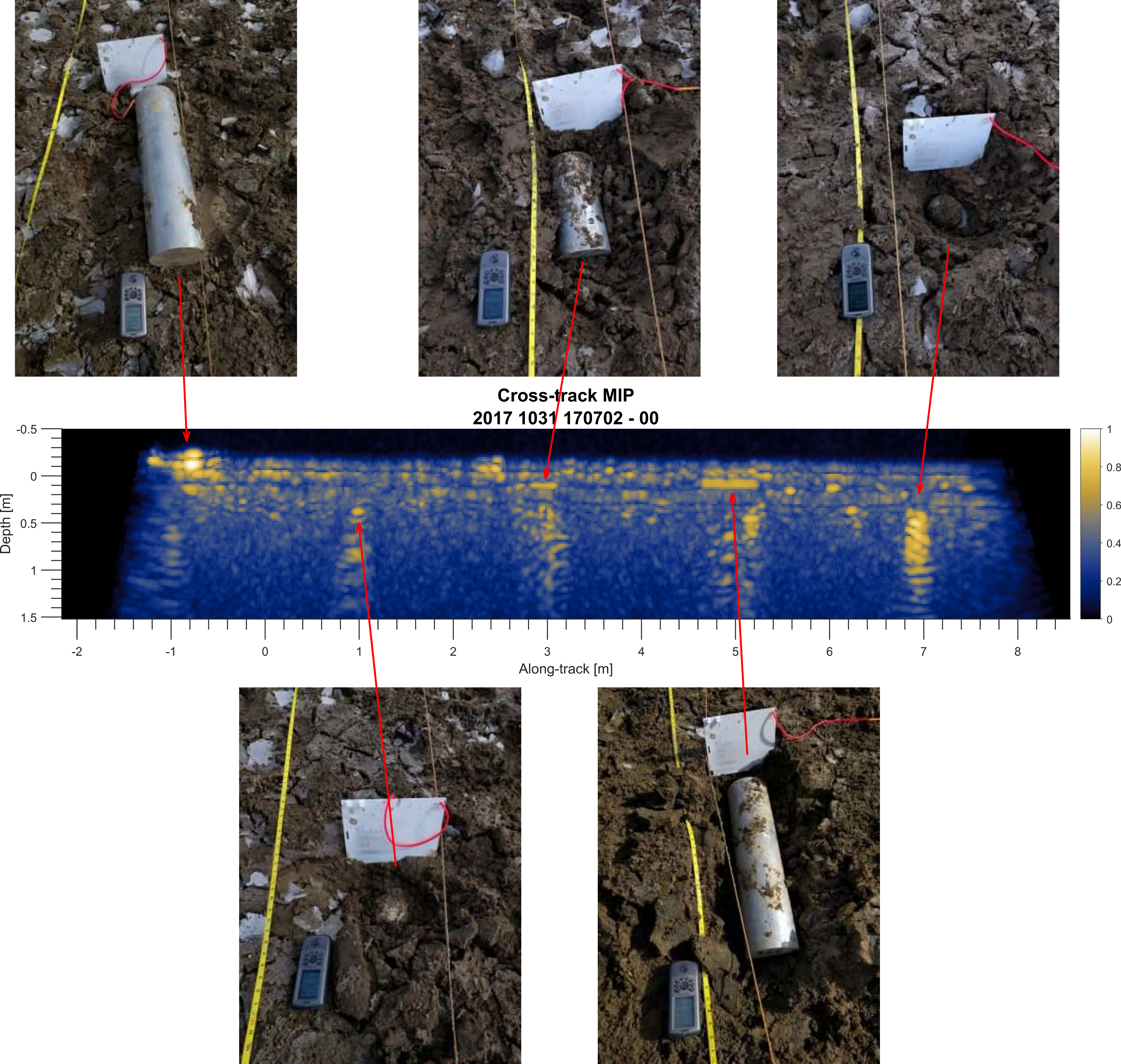}
  \caption{Five targets, S\_P09\_T3D0, S\_P08\_T1D1, S\_P07\_T2D1, S\_P06\_T3D1, and S\_P05\_T1D2, are shown in a cross-track MIP along with their installation photos.} \label{fig:targetsAndMip}
\end{figure*}

The survey data has also been processed to form maximum intensity projections (MIPs) along the three principal axes \cite{Wallis:1991a}. A cross-track MIP is shown in Figure~\ref{fig:targetsAndMip} after the data were preprocessed with background normalization and dynamic range compression algorithms. Targets were placed every \unit[2]{m} in the test field, and evidence of target return are seen at \unit[-1]{m}, \unit[1]{m}, \unit[3]{m}, \unit[5]{m}, and \unit[7]{m} along-track. The associated target installation photos are also shown in Figure~\ref{fig:targetsAndMip}. The large surface cylindrical target is distorted because it is truncated by the boundary of this image. Each of these targets shows a decaying return versus depth that extends far beyond the actual target dimension. It is hypothesized that these returns are due to elastic scattering phenomena. It is interesting that for the shot puts the level of the scattered elastic field appears to exceed that of the specular scattering. This result is a combination of two factors. First, in this location, the shot puts are placed near the silt/clay boundary which has a comparably strong acoustic reflection. This boundary reflection effectively masks the specular response of the shot put. The second factor is the result of the background normalization enhancing what is suspected to be the elastic response of the shot put. The result in this scenario is that the target return itself may not be directly observed in the imagery, yet the elastic response is clearly evident.

\section{Conclusion}
The problem of buried UXO detection is a current environmental problem facing the United States and other nations. Sensors do not exist to perform surveys in water depths less than \unit[5]{m}, and these depths are critical due to the potential for human/UXO interaction. Operation in these very shallow depths complicates the use of sub-bottom imaging systems hosted on either unmanned underwater vehicles or towed platforms. Additionally, the presence of interference from multipath reverberation can be challenging in these shallow depths.

To address these design challenges, a hybrid environmental scattering and target scattering model was used to study sensor performance across a range of environments and target types. This hybrid model combined PoSSM (environmental scattering) and TIER (target scattering) to create a model capable of producing realistic time series for a range of sensor designs, environments, and UXO. These modeling results were then used to inform a prototype sensor design.

Based on the modeling results, the prototype sensor was fielded on a \unit[9]{m} pontoon boat and tested at the Foster Joseph Sayers Reservoir in late 2017. Three different manmade objects as well as a few types of clutter targets were installed in the test area, across a range of water and sediment burial depths. In initial testing, the sensor was able to successfully image both proud and buried targets. The presence of elastic scattering phenomena are visible in numerous images collected by this sensor, and these scattering phenomena may provide a way to segment manmade targets from naturally occurring clutter \cite{Bucaro:2012a,Hall:2018a}.


\section*{Acknowledgment}
This research was supported in part by the U.S. Department of Defense, through the Strategic Environmental Research and Development Program (SERDP). The SERDP support was provided under the munitions response portfolio of Dr.~Herbert Nelson. This material is based upon work supported by the Humphreys Engineer Center Support Activity under Contract No. W912HQ-16-C-0006. This work was supported in part by the US Office of Naval Research contracts N00014-14-1-0539, N00014-14-1-0566, N00014-16-1-2313, and N00014-16-1-3022.

The authors were supported by numerous collaborators at ARL/PSU. We would like to thank Isaac Gerg and Steve Wagner for their efforts on the ASASIN beamformer, Norm Foster for supporting PoSSM development, and Zack Lowe for his work in developing and operating the sonar hardware on the test platform. The authors would also like to thank our collaborators at APL-UW, Aubrey Espa\~{n}a and Steve Kargl. Their work on the TIER model was critical to the modeling and simulation results. Finally, the authors thank Joe Calantoni and Ed Braithwaite for their support in sediment core analysis from the test site.




\bibliographystyle{./bib/IEEEtran}

\begin{thebibliography}{10}
\providecommand{\url}[1]{#1}
\csname url@rmstyle\endcsname
\providecommand{\newblock}{\relax}
\providecommand{\bibinfo}[2]{#2}
\providecommand\BIBentrySTDinterwordspacing{\spaceskip=0pt\relax}
\providecommand\BIBentryALTinterwordstretchfactor{4}
\providecommand\BIBentryALTinterwordspacing{\spaceskip=\fontdimen2\font plus
\BIBentryALTinterwordstretchfactor\fontdimen3\font minus
  \fontdimen4\font\relax}
\providecommand\BIBforeignlanguage[2]{{%
\expandafter\ifx\csname l@#1\endcsname\relax
\typeout{** WARNING: IEEEtran.bst: No hyphenation pattern has been}%
\typeout{** loaded for the language `#1'. Using the pattern for}%
\typeout{** the default language instead.}%
\else
\language=\csname l@#1\endcsname
\fi
#2}}

\bibitem{SERDP:2007a}
{Strategic Environmental Research and Development Program}, ``{SERDP} and
  {ESTCP} workshop on technology needs for the characterization, management,
  and remediation of military munitions in underwater environments,''
  \emph{SERDP-ESTCP Final Report}, 2007.

\bibitem{SERDP:2013a}
------, ``{SERDP}/{O}ffice of {N}aval {R}esearch workshop on acoustic detection
  and classification of {UXO} in the underwater environment,''
  \emph{SERDP-ESTCP Final Report}, 2013.

\bibitem{SERDP:2018a}
------, ``{SERDP} workshop on acoustic detection and classification of
  munitions in the underwater environment,'' \emph{SERDP-ESTCP Final Report},
  2018.

\bibitem{SERDP:2014a}
------, ``Informal workshop on burial and mobility modeling of munitions in the
  underwater environment,'' \emph{SERDP-ESTCP Final Report}, 2014.

\bibitem{Schock:2001a}
S.~Schock, A.~Tellier, J.~Wulf, J.~Sara, and M.~Ericksen, ``Buried object
  scanning sonar,'' \emph{{IEEE} J. Oceanic Eng.}, vol.~26, no.~4, pp.
  677--689, 2001.

\bibitem{Schock:2002a}
S.~G. Schock and J.~Wulf, ``Sonar for multi-aspect buried mine imaging,'' in
  \emph{MTS/IEEE OCEANS Conf.}, vol.~1, Oct 2002, pp. 479--484 vol.1.

\bibitem{Schock:2005a}
S.~G. Schock, J.~Wulf, G.~Quentin, and J.~Sara, ``Synthetic aperture processing
  of buried object scanning sonar data,'' in \emph{MTS/IEEE OCEANS Conf.}, Sept
  2005, pp. 2236--2241 Vol. 3.

\bibitem{Schock:2006a}
S.~G. Schock, J.~Wulf, and J.~Sara, ``Imaging performance of {BOSS} using {SAS}
  processing,'' in \emph{MTS/IEEE OCEANS Conf.}, Sept 2006, pp. 1--5.

\bibitem{Brown:2017b}
D.~C. Brown, S.~F. Johnson, and D.~R. Olson, ``A point-based scattering model
  for the incoherent component of the scattered field,'' \emph{J. Acoust. Soc.
  Am.}, vol. 141, no.~3, pp. EL210--EL215, 2017.

\bibitem{Kargl:2015a}
S.~G. Kargl, A.~L. España, K.~L. Williams, J.~L. Kennedy, and J.~L. Lopes,
  ``Scattering from objects at a water-sediment interface: Experiment,
  high-speed and high-fidelity models, and physical insight,'' \emph{IEEE J.
  Oceanic Eng.}, vol.~40, no.~3, pp. 632--642, July 2015.

\bibitem{Eckart:1953a}
C.~Eckart, ``The scattering of sound from the sea surface,'' \emph{J. Acoust.
  Soc. Am.}, vol.~25, no.~3, pp. 566--570, 1953.

\bibitem{APL-UW:1994a}
APL-UW, ``{APL-UW} high-frequency ocean environmental acoustic models
  handbook,'' Applied Physics Laboratory - University of Washington, Tech. Rep.
  TR 9407, Oct. 1994.

\bibitem{Urick:1983a}
R.~J. Urick, \emph{Principles of Underwater Sound}, 3rd~ed.\hskip 1em plus
  0.5em minus 0.4em\relax Los Altos, CA: Peninsula Publishing, 1983.

\bibitem{Hansen:2011a}
R.~E. Hansen, H.~J. Callow, T.~O. Sabo, and S.~A.~V. Synnes, ``Challenges in
  seafloor imaging and mapping with synthetic aperture sonar,'' \emph{IEEE
  Trans. Geosci. Remote Sensing}, vol.~49, no.~10, pp. 3677--3687, Oct 2011.

\bibitem{Lowe:2012a}
Z.~G. Lowe and D.~C. Brown, ``Multipath reverberation modeling for shallow
  water acoustics,'' in \emph{Proc. 11\textsuperscript{th} European Conference
  on Underwater Acoustics}, Edinburgh, Scotland, Jul. 2012, pp. 1285--1291.

\bibitem{Pierce:1991a}
A.~D. Pierce, \emph{Acoustics: {A}n introduction to its physical principles and
  applications}.\hskip 1em plus 0.5em minus 0.4em\relax Melville, NY:
  Acoustical Society of America, 1991.

\bibitem{PSDA:2018a}
\BIBentryALTinterwordspacing
{Pennsylvania Spatial Data Access}. (2018) {P}ennsylvania {I}magery
  {N}avigator. [Online]. Available: \url{http://www.pasda.psu.edu/}
\BIBentrySTDinterwordspacing

\bibitem{Dobeck:2010a}
G.~J. Dobeck, ``Adaptive large-scale clutter removal from imagery with
  application to high-resolution sonar imagery,'' vol. 7664, 2010, pp.
  76\,640X--76\,640X--10.

\bibitem{Williams:2018a}
D.~P. Williams, ``The {M}ondrian detection algorithm for sonar imagery,''
  \emph{IEEE Trans. Geosci. Remote Sensing}, vol.~56, no.~2, pp. 1091--1102,
  Feb 2018.

\bibitem{Wallis:1991a}
J.~W. Wallis and T.~R. Miller, ``Three-dimensional display in nuclear medicine
  and radiology.'' \emph{Journal of Nuclear Medicine}, vol.~32, no.~3, pp.
  534--546, 1991.

\bibitem{Bucaro:2012a}
J.~A. Bucaro, Z.~J. Waters, B.~H. Houston, H.~J. Simpson, A.~Sarkissian,
  S.~Dey, and T.~J. Yoder, ``Acoustic identification of buried underwater
  unexploded ordnance using a numerically trained classifier,'' \emph{J.
  Acoust. Soc. Am.}, vol. 132, no.~6, pp. 3614--3617, 2012.

\bibitem{Hall:2018a}
J.~J. Hall, M.~R. Azimi-Sadjadi, S.~G. Kargl, Y.~Zhao, and K.~L. Williams,
  ``Underwater unexploded ordnance {(UXO)} classification using a matched
  subspace classifier with adaptive dictionaries,'' \emph{IEEE J. Oceanic
  Eng.}, pp. 1--14, 2018.

\end{thebibliography}
%
%
%

\end{document}